\documentclass[pdflatex,sn-mathphys,iicol]{sn-jnl}

\usepackage{graphicx}%
\usepackage{multirow}%
\usepackage{amsmath,amssymb,amsfonts}%
\usepackage{amsthm}%
\usepackage{mathrsfs}%
\usepackage[title]{appendix}%
\usepackage{xcolor}%
\usepackage{textcomp}%
\usepackage{manyfoot}%
\usepackage{booktabs}%
\usepackage{algorithm}%
\usepackage{algorithmicx}%
\usepackage{algpseudocode}%
\usepackage{listings}
\usepackage{hyperref}
\usepackage{braket}
\usepackage[normalem]{ulem}
\usepackage{dsfont}
\usepackage{dcolumn}
\usepackage[square,numbers]{natbib}

\theoremstyle{thmstyleone}%
\theoremstyle{thmstyletwo}%

\theoremstyle{thmstylethree}%

\begin{document}

\title[Geometric phase gradient]{Adjoint computation of Berry phase gradients}


\author*[1,2]{\fnm{Cyrill} \sur{Bösch}}\email{cb7454@princeton.edu}

\author[2]{\fnm{Marc} \sur{Serra-Garcia}}\email{ M.SerraGarcia@amolf.nl}

\author[1]{\fnm{Christian} \sur{Böhm}}\email{christian.boehm@eaps.ethz.ch}

\author[1]{\fnm{Andreas} \sur{Fichtner}}\email{andreas.fichtner@eaps.ethz.ch}

\affil[1]{\orgdiv{Institue of Geophysics}, \orgname{ETH Zurich}, \orgaddress{\street{Sonneggstrasse 5}, \postcode{8092} \state{Zurich}, \country{Switzerland}}}

\affil[2]{\orgdiv{Department of Computer Science}, \orgname{Princeton University}, \orgaddress{\street{35 Olden Street}, \city{Princeton}, \state{NJ} \postcode{08540}, \country{US}}}

\affil[3]{\orgname{AMOLF}, \orgaddress{\street{Science Park 104}, \postcode{1098 XG} \state{Amsterdam}, \country{Netherlands}}}


\abstract{Berry phases offer a geometric perspective on wave propagation and are key to designing materials with topological wave transport. However, controlling Berry phases is challenging due to their dependence on global integrals over the Brillouin zone, making differentiation difficult. We present an adjoint-based method for efficiently computing the gradient of the Berry phase with respect to system parameters, involving only one forward and one adjoint calculation. This approach enables the use of advanced optimization techniques, such as topology optimization, to design new materials with tailored topological wave properties.}




\maketitle

\section{Introduction}\label{sec1}

The introduction of geometric phases to explain wave phenomena can be  attributed to Sir Michael Berry's groundbreaking paper, ``Quantal phase factors accompanying adiabatic changes" \cite{berry1984quantal}. These phases exhibit an inherent robustness as they only depend on the geometry of the path. Furthermore, in topological systems like the Su–Schrieffer–Heeger model, the Berry phase predicts the existence or absence of topological wave states. The field of topological physics was born with the observation and the explanation of the quantum Hall effect \cite{Klitzing80,TKNN}. Since then, geometric phases and topology have played a revolutionary role in multiple domains of physics, such as condensed matter physics \cite{PhysRevLett.107.186405, vanderbilt2018berry,tenfoldway,BerryPhaseCrystaline, neupert2018topological, LiangFuTopoCrystalline,Slager2013,kitaev2001unpaired}, topological photonic \cite{lu2014topological, ozawa2019topological, smirnova2020nonlinear,noh2018topological,rechtsman2013photonic,noh2018observation,rechtsman2013topological,miri2019exceptional,ni2023topological,peng2019probing, kang_topological_2023, zhang_non-abelian_2022, caceres-aravena_experimental_2022} and phononic metamaterials \cite{kumar2024role, huber2016topological,susstrunk2016classification,liu2018tunable, wang2015topological, riva2020edge, liu2017model, riva_adiabatic_2021, yan2018chip,vila2017observation, kane2014topological, prodan2009topological, berg2011topological, lubensky2015phonons, MatlackPerturb,SerraGarciaQuadrupole,periFragile, nassar2018quantization,Mousavi2015, liu2023classical, xia2021experimental}, quantum holonomic computing \cite{zanardi1999holonomic,solinas2004robustness,zhu2005geometric,fuentes2005holonomic}, the control and guiding of light \cite{jisha_geometric_2021, lin_light_2014, xu_inverse_2022, jisha_waveguiding_2023, escuti2016controlling, forbes2020structured} and seismology \cite{snieder2016seismic, boulanger2012observation}.

Recent developments in the field have shifted from fundamental research to exploiting these phenomena to invent new devices. For instance, quantum devices for signal processing, which leverage geometric phases as information carriers, have been introduced \cite{sjoqvist2015geometric,yin2009geometric,cusumano_geometric_2020,qiu2021experimental,lv2019accelerating}, topological states are harnessed to robustly propagate information \cite{shalaev2019robust, xue2021topological, yang2020terahertz, blanco2018topological}, and waveguides are designed based on the accumulation of a geometric phase as a wave packet is passed through a wave plate \cite{abbaszadeh_liquid-crystal-based_2021,jisha_waveguiding_2023,jisha_geometric_2021}. Of particular interest is the design of topological metamaterials \cite{zhu2018design, MatlackPerturb, SerraGarciaQuadrupole, MachineLearningPilozzi,PhononicOptimizationWaveguidingSR, PhononicOptimizationWaveguiding,FlorianML,TopoAcousticBandInversion,du_optimal_2020, wang_higher-order_2021, chen_inverse_2022,chen_inverse_2020, chen_inverse_2022} which is inherently challenging due to the intrinsic robustness of topological systems resulting from the fact that the topology of a material is invariant with respect to a large class of model perturbations. The design of these devices has predominantly been steered by intuition or top-down strategies, optimizing the desired output (e.g. the localization of energy at a specific location) rather than directly addressing the topological invariant.

In constrast, in our recent study \cite{bosch2020discovery}, we showed that topological invariants can be tuned using the concept of symmetry relaxation. This approach de-quantizes the invariant and allows a system to be systematically steered through a topological phase transition by following the gradient of the smoothed topological invariant with respect to design or control parameters. 

In this study,  we present an efficient method to compute gradients of Berry phases which often act as topological invariants in these systems. To this end, we employ the adjoint method \cite{long2015new,chachuat2007nonlinear,lee2007adjoint,hinze2008optimization,fichtner2010full} from the domain of PDE-constrained optimization. The adjoint method allows us to efficiently compute gradients, especially when the design space that characterises the device or material is high dimensional such as when optimizing continuous fields. 
The large design space renders the use of sampling algorithms or finite differentiation impractical. Such expansive design spaces are commonly encountered in the design of application devices, e.g., through topology optimization \cite{bendsoe2013topology}, contrasting with low-dimensional systems such as tight-binding or mass spring models.
Additionally, augmenting the number of design variables may allow us to identify designs with enhanced performance and robustness. The method introduced here can also be integrated in multi-objective optimization scenarios where, in addition to the Berry phase, one also optimizes, e.g., the quality factor or the operational bandwidth of topological states \cite{chen_inverse_2022}.

In a context broader than just topological invariants, the method is readily applicable to any single-band or multiband, non-Abelian geometric phase resulting from either cyclic or non-cyclic adiabatic evolutions, steered by either a Hermitian quantum-like first-order or a classical-like second-order dynamical system and variants that can be mapped to either of the two. 

In Section \ref{sec2}, we review the emergence of the Berry phase in both adiabatic quantum and classical systems, and state the gradient problem addressed in this article. In Section \ref{s:Adjoint method}, we derive the adjoint method, apply it to an elastic metamaterial crystal and provide a complexity analysis. Section \ref{s:Multiband} discusses the generalization to multiband Berry phases, derived from the Wilson loop eigenvalues. Finally, we conclude in Section \ref{s:Conclusions}.

\section{Preliminaries: quantum and classical Berry phases}\label{sec2}

We consider a discrete Hermitian non-autnomous quantum system whose dynamics are governed by a quantum Hermitian Hamiltonian, $\mathbf{H}(\mathbf{p}_H,t) \in \mathbb{C}^{N\times N}$, where $N$ is the state space dimension; or a classical system defined by a Hermitian mass matrix, $\mathbf{M}(\mathbf{p}_M)\in \mathbb{C}^{N\times N}$, and a Hermitian stiffness matrix, $\mathbf{K}(\mathbf{p}_K,t)\in \mathbb{C}^{N\times N}$. Without loss of generality, the $N_X$ parameters of the matrices, $\mathbf{p}_X \in \mathbb{R}^{N_X}$, are assumed to be real. The evolution of a quantum state $\mathbf{y}\in \mathbb{C}^N$ is given by the solution of the time-dependent Schrödinger equation
\begin{equation}
    i\partial_t\mathbf{y}(t) =  \mathbf{H}(\mathbf{p}_H,t)\mathbf{y}(t).
\end{equation}
The corresponding instantaneous eigenvalue problem is
\begin{equation}
    E_n(t)\mathbf{n}(t) =  \mathbf{H}(\mathbf{p}_H,t)\mathbf{n}(t),
\end{equation}
where $(E_n(t),\mathbf{n}(t))$ are the $n$-th instantaneous eigenvalue or energy and orthonormal eigenvector of $\mathbf{H}(\mathbf{p}_H,t)$. If $E_n(t)$ is non-degenerate for all integration times and the evolution of the Hamiltonian is slow enough in comparison to the energy gap, then a state initialized in the $n$-the eigenstate, i.e., $\mathbf{y}(0) = \mathbf{n}(0)$, remains in it and accumulates two types of phases \cite{kato1950adiabatic, sakurai1995modern, born1928beweis}:
\begin{equation}
  \mathbf{y}(t) \approx e^{-i\hat{E}_n(t)}e^{-i\theta_n(t)}\mathbf{n}(t),
\end{equation}
where $\hat{E}_n(t)$ is the dynamical phase, computed as
\begin{equation}
  \hat{E}_n(t) = \int_{0}^{t}E_n(t')dt',
\end{equation}
and $\theta_n(t)$ is the geometric phase, or Berry phase, which is obtained as 
\begin{equation}
  \theta_n(t) = \text{Im}{ \int_{0}^{t}\braket{\mathbf{n}(t')|\partial_t \mathbf{n}(t')}dt'}.
\end{equation}
The inner product is given by $\braket{\mathbf{a}|\mathbf{b}} = \sum_{i=1}^N a_i^*b_i$, and $*$ denotes complex conjugation.
For classical dynamics, the time-dependent equations of motion read
\begin{equation}
    \mathbf{M}(\mathbf{p}_M) \partial^2_t \mathbf{y}(t) =  \mathbf{K}(\mathbf{p}_K,t)\mathbf{y}(t).
\end{equation}
The corresponding generalized instantaneous eigenvalue problem is given by 
\begin{equation}\label{eq: classical eigval prob}
    \lambda_n(t)\mathbf{M}(\mathbf{p}_M)\mathbf{n}(t) = \mathbf{K}(\mathbf{p}_K,t)\mathbf{n}(t),
\end{equation}
where $\lambda_n(t) = \omega^2_n(t)$ is the frequency of oscillation squared. The eigenvectors are mass-orthonormal, i.e., $\braket{\mathbf{n}|\mathbf{M}|\mathbf{m}} = \delta_{nm}$. If the system is initialized as $\mathbf{y}(0) = \mathbf{n}(0)$ and evolved slow enough, then the solution is well described by \cite{nassar2018quantization, bosch2023differences}
\begin{equation}
    \mathbf{y}(t) \approx \sqrt{ \frac{\omega_n(0)}{\omega_n(t)}}  e^{-    i\Omega_n(t)}e^{-i\gamma_n(t)}\mathbf{n}(t) ,
\end{equation}
where
\begin{equation}
  \Omega(t) = \int_{0}^{t}\omega_n(t')dt',
\end{equation}
and the classical geometric phase reads
\begin{equation}\label{eq:Berry phase continuous}
  \gamma_n(t) = \text{Im}{ \int_{0}^{t}\braket{\mathbf{n}(t')|\mathbf{M}(\mathbf{p}_M)|\partial_t \mathbf{n}(t')}dt'}.
\end{equation}
A detailed discussion on the differences between quantum and classical adiabatic evolution and an analysis under which respective conditions the adiabatic approximation is valid, can be found in \cite{bosch2023differences}.

We assume that the mass and in particular the stiffness matrix are defined over the complex field to cover, e.g., Bloch systems, as discussed in Section \ref{s:Application}. This implies that the state vector $\mathbf{y}$ can be complex. Real-valued systems can, however, also have a non-trivial Berry phase \cite{bosch2023differences, kariyado2016hannay}. Here, the Berry phase encodes the information of whether or not the parallel transported eigenstate reaches the opposite side of the real $N$-unit sphere or returns to its initial position after a closed loop evolution.

In what follows, we will constrain ourselves to the classical case and note that the quantum Berry phase $\theta_n$ is a special case of the classical Berry phase $\gamma_n$ by choosing $\mathbf{M}(\mathbf{p}_M) := \mathds{1}$ and $\mathbf{H}(\mathbf{p}_H,t) := \mathbf{K}(\mathbf{p}_K,t)$. 

The Berry phase is quantized to $0$ or $\pi$ modulo $2\pi$ and acts as a topological invariant in the presence of protecting symmetries \cite{neupert2018topological, vanderbilt2018berry}. Relaxing those symmetries dequantizes the Berry phase and allows us to compute derivatives \cite{bosch2020discovery}. In case of a real-valued system, the Berry phase is protected by the absence of imaginary (self-) couplings. Therefore, differentiability for such a system requires a complex parametrization that can interpolate the two real systems with Berry phases $0$ or $\pi$.

The contribution of this article is to derive the adjoint method to efficiently compute this parameter derivative $d\gamma_n(t)/d\mathbf{p}_X$. Furthermore, the gradient is exact within the numerical discretization error of the forward problem. This allows us to systematically stear a system through topological phase transition following updates of the form $\mathbf{p}_X \rightarrow \mathbf{p}_X - \alpha d\gamma_n^c/d\mathbf{p}_X$, where $\alpha$ is some optimal step length. Furthermore, the gradient informs us about the robustness of geometric phase devices beyond symmetric perturbation.

\section{Adjoint method for geometric phase derivatives} \label{s:Adjoint method}

To compute the gradient, we will employ an approximate expression of the Berry phase on the discretized interval $t \in [0,t_I]$ with discrete time points $\{t_i\}_{i=0}^I$ \cite{vanderbilt2018berry}:
\begin{multline}\label{eq: Berry phase formula used for adjoint}
\gamma_n \big[\mathbf{n}(t_0),...,\mathbf{n}(t_{I}),\mathbf{p}_M\big] \\ \approx  \sum_{i=0}^{I} \text{Im}\ln \Braket{\mathbf{n}(t_i)|\mathbf{M}(\mathbf{p}_M)| \mathbf{n}(t_{i+1})},
\end{multline}
where, in the interest of the following derivation, we made the eigenvector and parameter dependence of the Berry phase explicit. In the limit where $I \rightarrow \infty$ the above expression converges to the exact Berry phase in Equation \eqref{eq:Berry phase continuous}.
We further assume that the loop is closed, i.e. $\mathbf{n}(t_0) = \mathbf{n}(t_I)$, which implies that the Berry phase is gauge invariant, as each eigenvector appears once as a bra and once as a ket. The complex conjugation therefore cancels any phase change in the individual eigenvectors \cite{vanderbilt2018berry}. 
Consequently, the gradient derived from it also inherits gauge invariance. Equation \eqref{eq: Berry phase formula used for adjoint} proves to be the simplest and most practical for our intended purposes. Notably, any standard eigenvalue solver can be used to determine the eigenstates along the trajectory without necessitating special consideration for relative phase handling. Additionally, this approach possesses the advantage of emulating exact adiabatic evolution even with sparsely sampled time points, which, in our experience, is computationally more efficient than direct adiabatic integration of the time-dependent problem.   

Note that equation \eqref{eq: Berry phase formula used for adjoint} covers also the case of a non-cyclic Berry phase where the loop is closed along any geodesic that connects $\mathbf{n}(t_0) $ and $ \mathbf{n}(t_{I-1})$ \cite{samuel1988general, leone2019parallel, pancharatnam1956generalized}.

We simplify the notation by introducing $\mathbf{p} := [\mathbf{p}_M,\mathbf{p}_K] \in \mathbb{R}^{N_p}$
and defining $\mathbf{M} := \mathbf{M}(\mathbf{p})$ for the time-dependent mass matrix, $\mathbf{K}_i := \mathbf{K}(\mathbf{p},t_i)$ for the stiffness matrix, $\lambda_i := \lambda_n(t_i)$ for the instantaneous eigenvalues, and $\mathbf{n}_i := \mathbf{n}(t_i)$ for the eigenvectors.

\subsection{From constrained to unconstrained}
In practice, the Berry phase of the system will be optimized to some target value $\gamma^{tar.}$, i.e. we minimize an objective function of the form $J =1/2 ||\gamma_n -\gamma^{tar.}||^2 $. Chain rule implies that $dJ/d\mathbf{p} = \partial J/\partial \gamma_n d\gamma_n/d\mathbf{p} = (\gamma_n -\gamma^{tar.}) d\gamma_n/d\mathbf{p}$. To simplify the notation of the following gradient derivation we cast the optimization problem in terms of minimizing $\gamma_n$ rather than $J$.

Direct computation of $d\gamma_n^c/d\mathbf{p} = d\gamma_n^c/d\mathbf{n}_i \partial \mathbf{n}_i/\partial\mathbf{p}$ entails evaluating $\partial \mathbf{n}_i/\partial\mathbf{p}$. However, computing and storing the $\mathbb{C}^{N\times N_p}$ matrix $\partial \mathbf{n}_i/\partial\mathbf{p}$ becomes impractical when $\mathbf{n}$ and $\mathbf{p}$ are high-dimensional. The adjoint procedure presents a solution to eliminate the need for $\partial \mathbf{n}_i/\partial\mathbf{p}$ in gradient computation. 

Following the developments in \cite{hinze2008optimization} and \cite{lee2007adjoint}, we address the constrained optimization problem:
\begin{multline}\label{eq:constrained single band opt problem}
\min_{\mathbf{n}_0,...,\mathbf{n}_{I},\mathbf{p}} \gamma_n\big[\mathbf{n}_0,...,\mathbf{n}_{I},\mathbf{p}\big] 
\quad \textrm{s.t.} \quad 
\\ \left(\mathbf{K}_i-\lambda_i \mathbf{M}\right) \mathbf{n}_i = 0, 
\\1-\braket{\mathbf{n}_i|\mathbf{M}| \mathbf{n}_i} = 0 \quad i = 1,..., I , 
\end{multline}
The first $I$ constraints ensure that the pairs $(\lambda_i,\mathbf{n}_i)$ are solutions to the eigenvalue problem \eqref{eq: classical eigval prob}. The subsequent set of $I$ constraints enforce that the eigenvectors lie on the mass-induced unit sphere, ensuring well-posedness of the system of equations. We assume that a unique solution (up to a phase) exists for the eigenvalue problem, denoting it as $(\lambda_i,\mathbf{n}_i)= ( \lambda_i (\mathbf{p}),\mathbf{n}_i(\mathbf{p}))$ for a particular $\mathbf{p}$. Further, we assume differentiability of $\mathbf{K}_i$ and $\mathbf{M}$ with respect to $\mathbf{p}$.

The transformation of the constrained optimization problem into an unconstrained one is achieved through the concept of Lagrange multipliers. In the realm of PDE-constrained optimization, the Lagrange multipliers correspond to adjoint variables, as will become evident. The Lagrangian takes the form
\begin{multline}\label{eq: langragian}
\mathcal{L}\big[\mathbf{n}_0,...,\mathbf{n}_{I},\lambda_0,...,\lambda_{I},\mathbf{p},\mathbf{u}_0,...,\mathbf{u}_{I},v_0,...,v_{I}\big] = \\
\gamma_n\big[\mathbf{n}_0,....,\mathbf{n}_{I}, \mathbf{p}\big] 
\\ +\sum_{i=0}^{I}  \Big\langle \mathbf{u}_i| \mathbf{K}_i-\lambda_i \mathbf{M}| \mathbf{n}_i \Big\rangle 
\\ + \sum_{i=0}^{I}\frac{1}{2}  v_i^*\Big(1-\braket{\mathbf{n}_i|\mathbf{M}| \mathbf{n}_i}\Big),
\end{multline}
where $\mathbf{u}_i \in \mathbb{C}^n$ and $v_i \in \mathbb{C}$ represent the adjoint variables. The Lagrangian can be evaluated for any $(\lambda_i,\mathbf{n}_i)$. Selecting $(\lambda_i,\mathbf{n}_i)$ to be a solution of the eigenvalue problem, i.e., $(\lambda_i,\mathbf{n}_i) = ( \lambda_i (\mathbf{p}),\mathbf{n}_i(\mathbf{p}))$, renders the sum in the equation above equal to $0$. We denote the Langragian evaluated for this particular choice of $\mathbf{p}$ as $\mathcal{L}|_\mathbf{p}$ and similarly the geometric phase as $\gamma_n|_\mathbf{p}$. We find that 
\begin{equation}
 \mathcal{L}|_\mathbf{p} = \gamma_n|_\mathbf{p}  + 0 
 = \gamma_n|_\mathbf{p}.
 \end{equation} 
 Consequently, taking the design parameter derivative at the points $(\mathbf{n}_i, \lambda_i )= (\mathbf{n}_i(\mathbf{p}), \lambda_i (\mathbf{p}))$, we obtain
\begin{equation} \label{eq: Berry phase derivative full}
   \frac{d\gamma_n|_\mathbf{p}}{d p_m} = \sum_{i = 0}^{I} \partial_{\mathbf{n}_i} \mathcal{L}|_\mathbf{p} \partial_{p_m} \mathbf{n}_i
   + \partial_{p_m}  \mathcal{L}|_\mathbf{p} + \partial_{p_m}  \gamma_n|_\mathbf{p},
\end{equation}
where
\begin{equation}
    \partial_{p_m}  \gamma_n|_\mathbf{p} = \sum_{i=1}^{I} \text{Im}\ln \Braket{\mathbf{n}_i(\mathbf{p})|\partial_{p_m}\mathbf{M}| \mathbf{n}_{i+1}(\mathbf{p})}.
\end{equation}

\subsection{The adjoint problem and the gradient}
Among all possible choices for $\mathbf{u}_0 ,...,\mathbf{u}_{I}$ and $v_0,...,v_{I}$, let us denote $\mathbf{u}_0(\mathbf{p}),..., \mathbf{u}_{I}(\mathbf{p})$ and $ v_0(\mathbf{p}),...,v_{I}(\mathbf{p})$ as the adjoint variables that satisfy $\partial \mathcal{L}|_{\partial{p}}/\partial \mathbf{n}_i = 0$. When such adjoint variables exist, the first term in the above sum becomes zero. The equations emerging from equation \eqref{eq: langragian} when requireing that $\partial \mathcal{L}|_{\partial{p}}/\partial \mathbf{n}_i = 0$ are referred to as the adjoint problem, given by
\begin{multline} \label{eq: Adjoint Problem}
   \begin{bmatrix}
      \mathbf{K}_i-\lambda_i \mathbf{M} & -\mathbf{M} \mathbf{n}_i(\mathbf{p})\\
      -\mathbf{n}_i^{\dagger}(\mathbf{p})\mathbf{M} & 0 
   \end{bmatrix}
   \begin{bmatrix}
      \mathbf{u}_i(\mathbf{p})\\ v_i(\mathbf{p})
   \end{bmatrix}
   \\=
   \begin{bmatrix}
      -(\partial_{\mathbf{n}_i} \gamma_n)^\dagger \\ 
      0
   \end{bmatrix}.
\end{multline}
The superscript $\dagger$ denotes conjugate transposition. The right-hand side is the adjoint source and can be obtained analytically as demonstrated below. The matrix on the left-hand side is proven to be invertible in \cite{lee2007adjoint}. Hence,  $\mathbf{u}_0(\mathbf{p}),..., \mathbf{u}_{I}(\mathbf{p})$ and $  v_0(\mathbf{p}),...,v_{I}(\mathbf{p})$ exist and are unique. Therefore, with the choice of  $\mathbf{u}_0 = \mathbf{u}_0(\mathbf{p}),...,\mathbf{u}_{I} = \mathbf{u}_{I}(\mathbf{p})$ and $v_0 =  v_0(\mathbf{p}),...,v_{I}= v_{I}(\mathbf{p})$, we have $\partial \mathcal{L}/\partial \mathbf{n}_i = 0$, and equation \eqref{eq: Berry phase derivative full} reduces to
\begin{multline}\label{eq:design parameter gradient}
   \frac{d\gamma_n|_{\mathbf{p}}}{d p_m} = \partial_{p_m} \mathcal{L}|_{\mathbf{p}} +\partial_{p_m} \gamma_n|_\mathbf{p}
   \\ = \sum_{i = 0}^{I} \Big\langle \mathbf{u}_i(\mathbf{p})|\partial_{p_m}\mathbf{K}_i - \lambda_i \partial_{p_m} \mathbf{M}|\mathbf{n}_i(\mathbf{p}) \Big\rangle
   \\ +\sum_{i = 0}^{I}  \frac{1}{2} v_i^*(\mathbf{p})\Big(1-\braket{\mathbf{n}_i(\mathbf{p})|\partial_{p_m}\mathbf{M}| \mathbf{n}_i(\mathbf{p})}\Big) \\+\sum_{i = 0}^{I} \text{Im}\ln \Braket{\mathbf{n}_i(\mathbf{p})|\partial_{p_m}\mathbf{M}| \mathbf{n}_i(\mathbf{p})} ,
\end{multline}
where the design parameter derivative of the stiffness and mass matrices can often be obtained analytically. Note that the Berry phase and hence its gradient only exist for parameters where the eigenvalue gap to other eigenstates does not vanish. Otherwise, the Berry phase is not well-defined. It remains to compute the adjoint source. 

\subsection{The adjoint source}
To avoid considerations about the analyticity of the complex eigenstates involved, we treat the real and imaginary parts as separate components of a vector for the differentiation of the Berry phase. 
The augmented state vectors are defined as
\begin{equation} \label{eq: aug state}
\hat{\mathbf{n}}_i = \begin{bmatrix} \text{Re} \, \mathbf{n}_i \\ \text{Im} \, \mathbf{n}_i \end{bmatrix}^T \in \mathbb{R}^{2N },
\end{equation}
and the augmented symmetric mass matrix
\begin{equation}\label{eq: aug mass}
    \hat{\mathbf{M}} = \begin{bmatrix}
        \mathbf{M} & 0 \\ 0 & \mathbf{M}
    \end{bmatrix}.
\end{equation}
We further define 
\begin{equation}\label{eq: Q}
\hat{\mathbf{Q}} = \begin{bmatrix} \mathbf{0}_{N \times N } & \mathds{1}_{N \times N }\\ -\mathds{1}_{N \times N } & \mathbf{0}_{N \times N } \end{bmatrix} \in \mathbb{R}^{2N \times 2N}.
\end{equation}
In this augmented space, the original mass induced inner product, $\braket{\mathbf{n}|\mathbf{M}|\mathbf{m}}$, can be computed as $\braket{\mathbf{n}|\mathbf{M}|\mathbf{m}} = \braket{\hat{\mathbf{n}}|\hat{\mathbf{M}}|\hat{\mathbf{m}}} + i \braket{\hat{\mathbf{n}}|\hat{\mathbf{Q}}\hat{\mathbf{M}}|\hat{\mathbf{m}}}$. We denote the local phase mismatches or local Berry phases, where $\mathbf{n}_i$ contributes, as 
\begin{multline}
    \gamma_n^{i-1} = \text{Im}\ln \Braket{\mathbf{n}_{i-1}|\mathbf{M}| \mathbf{n}_{i}} \\ =\text{Im}\ln \big( \braket{\hat{\mathbf{n}}_{i-1}|\hat{\mathbf{M}}|\hat{\mathbf{n}}_{i}} + i \braket{\hat{\mathbf{n}}_{i-1}|\hat{\mathbf{Q}}\hat{\mathbf{M}}|\hat{\mathbf{n}}_{i}}\big),
\end{multline}
and 
\begin{multline}
    \gamma_n^i = \text{Im}\ln \Braket{\mathbf{n}_i|\mathbf{M}| \mathbf{n}_{i+1}}  \\ =\text{Im}\ln \big( \braket{\hat{\mathbf{n}}_i|\hat{\mathbf{M}}|\hat{\mathbf{n}}_{i+1}} + i \braket{\hat{\mathbf{n}}_i|\hat{\mathbf{Q}}\hat{\mathbf{M}}|\hat{\mathbf{n}}_{i+1}}\big),
\end{multline}
where the order in the inner product is swapped for $\gamma_n^{i-1}$. Since $\mathbf{n}_i$ only appears in these two terms of the Berry phase, we have
\begin{equation}
\partial_{\hat{\mathbf{n}}_i}\gamma_n = \partial_{\hat{\mathbf{n}}_i}\gamma_n^{i-1} + \partial_{\hat{\mathbf{n}}_i}\gamma_n^i \in \mathbb{R}^{1 \times 2N}.
\end{equation}
Using $\partial_x \ln x = 1/x$, we can compute the partial derivatives of the two contributions in the augmented space as
\begin{multline}
\partial_{\hat{\mathbf{n}}_i}\gamma_n^{i-1} = \frac{1}{\braket{\hat{\mathbf{n}}_i|\hat{\mathbf{M}}|\hat{\mathbf{n}}_{i-1}}^2 + \braket{\hat{\mathbf{n}}_i|\hat{\mathbf{Q}}\hat{\mathbf{M}}|\hat{\mathbf{n}}_{i-1}}^2} 
\\ \Big( \braket{\hat{\mathbf{n}}_i|\hat{\mathbf{M}}|\hat{\mathbf{n}}_{i-1}}\hat{\mathbf{n}}^T_{i-1}\hat{\mathbf{Q}}\hat{\mathbf{M}}  - \\ \braket{\hat{\mathbf{n}}_i|\hat{\mathbf{Q}}\hat{\mathbf{M}}|\hat{\mathbf{n}}_{i-1}}\hat{\mathbf{n}}^T_{i-1}\hat{\mathbf{M}}\Big)
\end{multline}
and
\begin{multline}
\partial_{\hat{\mathbf{n}}_i}\gamma_n^i = \frac{1}{\braket{\hat{\mathbf{n}}_i|\hat{\mathbf{M}}|\hat{\mathbf{n}}_{i+1}}^2 + \braket{\hat{\mathbf{n}}_i|\hat{\mathbf{Q}}\hat{\mathbf{M}}|\hat{\mathbf{n}}_{i+1}}^2} 
\\ \Big( \braket{\hat{\mathbf{n}}_i|\hat{\mathbf{M}}|\hat{\mathbf{n}}_{i+1}}\hat{\mathbf{n}}^T_{i+1}\hat{\mathbf{M}}\hat{\mathbf{Q}}^T  - \\ \braket{\hat{\mathbf{n}}_i|\hat{\mathbf{Q}}\hat{\mathbf{M}}|\hat{\mathbf{n}}_{i+1}}\hat{\mathbf{n}}^T_{i+1}\hat{\mathbf{M}}\Big).
\end{multline}
With these definitions, the adjoint source at time $t = t_i$ can be found as
\begin{multline}\label{app eq: adjoint source}
\partial_{\mathbf{n}_i}\gamma_n|_{\mathbf{p}} = \partial_{\hat{\mathbf{n}}_i}\gamma_n[1:N] \\+ i\partial_{\hat{\mathbf{n}}_i}\gamma_n[N+1:2N] \in \mathbb{C}^{1 \times N}.
\end{multline}

\subsection{The algorithm}
In summary, to compute the design parameter gradient, we follow Algorithm \ref{algo}.
\begin{algorithm}\label{algo}
\caption{Berry phase gradient}\label{alg:cap}
\begin{algorithmic}
\State 1. Solve $\lambda_i \mathbf{M}(t_i) \mathbf{n}_i= \mathbf{K}(t_i)\mathbf{n}_i$ for $\{t_i\}_{i=0,...,I}$ to obtain $\{\mathbf{n}_i\}_{i=0,...,I}$,  $\{\lambda_i\}_{i=0,...,I}$ 

\State 2. Compute the adjoint sources $\partial_{\mathbf{n}_i} \gamma_n|_\mathbf{p}$ for $\{t_i\}_{i=0,...,I}$

\State 3. Solve adjoint problem \eqref{eq: Adjoint Problem} to obtain adjoint variables  $\{\mathbf{u}_i(\mathbf{p})\}_{i=0,...,I}$ and  $\{v_i(\mathbf{p})\}_{i=0,...,I}$

\State 4. Evaluate $\partial_{p_m} \mathbf{M}$ and $ \partial_{p_m}\mathbf{K}_i$ 

\State 5. Compute the design parameter gradient as in equation \eqref{eq:design parameter gradient} for all model parameters $p_0,...,p_{N_p}$

\end{algorithmic}
\end{algorithm}
We emphasize that if $\partial_{p_m} \mathbf{K}_i$ and $\partial_{p_m} \mathbf{M}$ can be obtained analytically, they do not need to be explicitly constructed. Instead, equation \eqref{eq:design parameter gradient} can be efficiently implemented as a vector-matrix product and steps 4 and 5 of Algorithm \ref{algo} can be merged. This approach significantly reduces computational overhead and memory requirements.

In the context of optimizing the geometric phase, it is necessary to maintain a gap between adjacent eigenvalue levels. Remarkably, within this framework, the design parameter gradient of the gap can be derived from the gradient of the eigenvalues, which is readily computed as \cite{lee2007adjoint}
\begin{equation}
   \frac{d\lambda_i}{d p_m} = \Big\langle \mathbf{n}_i(\mathbf{p})|\partial_{p_m} \mathbf{K}_i - \lambda_i \partial_{p_m} \mathbf{M}|\mathbf{n}_i(p) \Big\rangle.
\end{equation}

\subsection{Application to a continuous 1D elastic topological metamaterial: Hockey-Stick-Test}\label{s:Application} 
\begin{figure*}
   \makebox[\textwidth][c]{\includegraphics[width=1.2\linewidth]{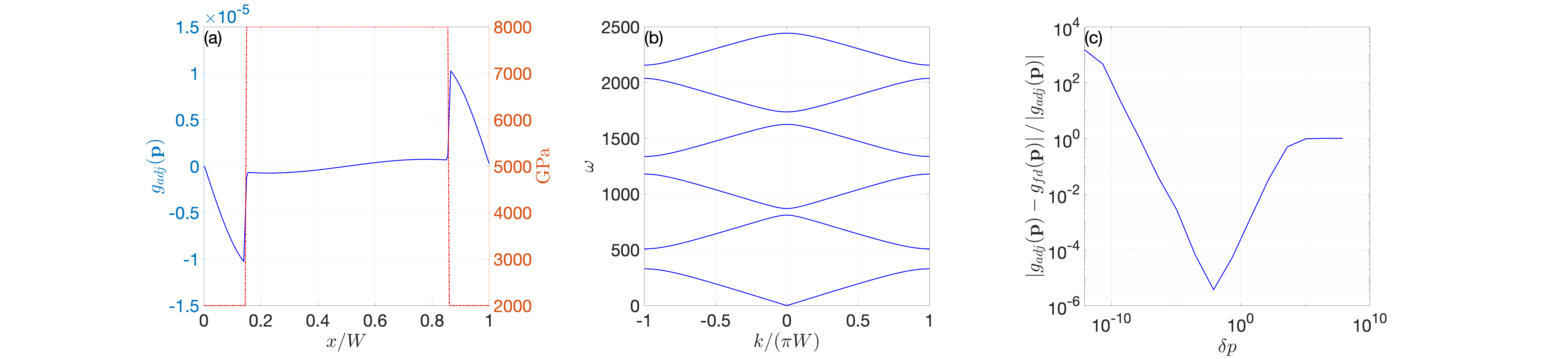}}
   \caption{(a) Elastic modulus profile across the unit cell in red and the gradient of the Berry phase of the 1st band in blue. From the profile, it is evident that altering the Zak phase necessitates breaking its inversion symmetry. (b) Bloch band structure over the Brillouin Zone. (c) Hockey-Stick-Test: The y-axis depicts the relative disparity between the adjoint and first-order finite differentiation  gradients across varying design parameter perturbations (x-axis). The distinctive ``hockey stick'' curvature of the plot robustly underscores the accuracy of the adjoint gradient.}
   \label{fig1}
 \end{figure*}

In this section, we utilize the adjoint method to calculate the gradient of the Berry phase of a continuous elastic metamaterial rod. To validate the correctness of the gradient, we subject it to a Hockey-Stick-Test \cite{fichtner2021lecture}. We consider a one-dimensional elastic crystal, which, when inversion symmetry is present, can exhibit a topological phase \cite{tsai2019topological, xiao2014surface}. This elastic rod is aligned along the $x$-axis and has a unit cell of width $W$, with elastic modulus $E(x + W) = E(x)$ and density $\rho(x + W) = \rho(x)$. The rod satisfies the linear elastic wave equation
\begin{equation}\label{eq:ElasticRodEQofMotion}
\rho(x) \partial^2_t y(t,x) = \partial_x \Big[\text{E}(x) \partial_x y(t,x)\Big],
\end{equation}
where $y$ represents the transverse or horizontal displacement along the longitudinal direction. The generalized eigenvalue problem reads
\begin{equation}\label{eq:generalized Bloch eigval problem}
\lambda_n(k) \rho(x) n(x,k) =  \partial_x \text{E}(x) \partial_x n(x,k).
\end{equation}
The Bloch theorem \cite{ashcroft1976solid, vanderbilt2018berry} guarantees that the eigenvectors are given by $y_n(x,k) = \exp(-i kx/W)n(x,k)$, where $k$ is the Bloch wavenumber. Here, $n(x,k)$ corresponds to the Bloch wave functions, exhibiting the same periodicity as the lattice, i.e., $n(x+W,k) = n(x,k)$. The corresponding instantaneous frequency, $\omega_n(k)$, is given by the square root of the eigenvalue $\lambda_n(k)$. Plotting $\lambda_n(k)$ or $\omega_n(k)$ for each $n$ as a function of $k$ yields the band structure, as shown in Figure \ref{fig1}(b). The eigenvalues and Bloch wave functions are determined through the generalized Bloch eigenvalue problem
\begin{multline}\label{eq:generalized Bloch eigval problem}
\lambda_n(k) \rho(x) n(x,k)  \\= e^{ikx/W}\Big[ \partial_x \text{E}(x) \partial_x\Big]e^{-ikx/W}n(x,k)
\\ =  \Bigl[\partial_x E(x) \partial_x  +i k \Big(2E(x)\partial_x + \partial_x E(x)\Big) \\ - E(x)k^2\Bigr]n(x,k).
\end{multline}
Now, we employ the spectral-element method \cite{fichtner2010full,igel2017computational} to discretize this eigenvalue problem. The spectral-element method is a finite-element-like discretization scheme that produces diagonal mass matrices and has been recognized for its effectiveness in solving wave problems.
 We define the $N \times N$ assembly stiffness matrices obtained by discretizing the above operators as $\partial_x E(x) \partial_x \rightarrow \mathbf{K}_a$, $2E(x)\partial_x+ \partial_x E(x) \rightarrow \mathbf{K}_b$, and $E(x) \rightarrow \mathbf{K}_c$. The discretized Bloch stiffness matrix is given by
\begin{equation}
   \mathbf{K}\big(\mathbf{p},k\big) = \frac{1}{\rho}\big( \mathbf{K}_a(\mathbf{p}) + i k \mathbf{K}_b(\mathbf{p}) - k^2\mathbf{K}_c(\mathbf{p})\big).
\end{equation}
For simplicity, we assume a constant value for the density and absorb it into the stiffness matrix. However, the method can be straightforwardly generalized to accommodate arbitrary density profiles. Finally, discretizing $n \rightarrow \mathbf{n}$ we obtain the generalized eigenvalue problem \eqref{eq: classical eigval prob} for a constant mass matrix and where the Bloch wavenumber plays the role of time.

It can be shown that the eigenvalues are periodic in $k$-space, meaning that $\lambda_n(k+2\pi/W) = \lambda_n(k)$ \cite{ashcroft1976solid}. From now on, we will work in a periodic gauge \cite{vanderbilt2018berry}, meaning that we have $\mathbf{n}(k+2\pi/W) = \mathbf{n}(k)$ (not just up to a phase). This implies that we can confine ourselves to the Broullion Zone (BZ), which ranges from $-\pi/W$ to $\pi/W$. The periodic gauge relates Bloch functions as follows: $\exp(-i (k+2\pi)x/W)\mathbf{n}(k+2\pi) = \exp(-i kx/W)\mathbf{n}(k)$, which implies that $\mathbf{n}(k)=\exp(-i kx/W)\mathbf{n}(k+2\pi)$. Let us divide the BZ into discrete intervals of Bloch wavenumber $k_0 = -\pi/W,....,k_{I} = \pi/W$. While we opt for equidistantly distributed $k$-points in this analysis, this choice is not mandatory.
The single-band Berry phase, known as the Zak phase in this context, can be computed as
\begin{multline}\label{eq: Elastic Rod Berry Phase}
   \gamma_n = \sum_{i=0}^{N-1} \text{Im} \ln \braket{\mathbf{n}(k_i)|\mathbf{n}(k_{i+1})} \\ + \text{Im} \ln \braket{\mathbf{n}(k_{I})|e^{-i 2\pi x/W}|\mathbf{n}(k_0)}.
\end{multline}
The additional phase factor, $\exp(-i kx/W)$, that emerges is a peculiar characteristic of the Zak phase \cite{zak1989berry, vanderbilt2018berry}. In systems with, e.g., inversion symmetry, the Zak phase is quantized to $0$ or $\pi$ (modulo $2\pi$) \cite{zak1989berry,xiao2014surface,neupert2018topological}. However, in systems lacking symmetries, the Berry phase can assume any value, and it exhibits almost everywhere differentiability \cite{bosch2020discovery}. If discretized, the Zak phase may act as a topological invariant, which, through the Bulk-Edge-Correspondence, predicts the existence or absence of topological surface modes.

We now turn our attention to computing the gradient of the Zak phase for the first ($n=0$) band with respect to all model parameters, denoted as $d\gamma_0/d\mathbf{p}$. In the specific example presented here, we employed $N=200$ grid points, and an equal number of model parameters, i.e., $N_p = N$. The gradient obtained from the adjoint method is represented by $\mathbf{g}_{adj}(\mathbf{p}) = d\gamma_0/d\mathbf{p} \in \mathbb{R}^{N_p}$. We evaluate this gradient for the metamaterial rod. The rod's elastic modulus profile $E(x)$ is depicted Figure \ref{fig1}(a) and the density is set to a constant value of $\rho =  2704\ \text{kg/m}^3$. In this configuration, we find that $\gamma_0(\mathbf{p}) = 0$, indicating a topologically trivial phase. The Zak phase is quantized, a result of the inversion symmetry. In Figure \ref{fig1}(a), the gradient distribution throughout the unit cell is depicted on top of the elastic modulus profile. The gradient informs us that we have to break the inversion symmetry to change the Zak phase.

We aim to compare the gradient derived from the adjoint method, $\mathbf{g}_{adj}$, with the gradient obtained through first-order finite differentiation, $\mathbf{g}_{fd}$ through a Hockey-Stick test \cite{fichtner2021lecture}. The latter is determined by perturbing each model parameter individually by a small amount, $\delta p$, and subsequently recalculating all the eigenvectors along the path. Specifically, for a perturbation in the $m$-th model parameter, the perturbed parameter set is $\mathbf{p}_m^{pert} = [p_1,...,p_m+\delta p,...,p_{\mathcal{M}}]$. The gradient through first-order finite differentiation for each model parameter (here equivalently, for each numerical grid point) is then given by 
\begin{equation}
[\mathbf{g}_{fd}(\mathbf{p})]_m = \frac{\gamma_0(\mathbf{p}_m^{pert})-\gamma_0(\mathbf{p})}{\delta p}.
\end{equation} 
The Hockey-Stick-Test involves successively reducing $\delta p$ and juxtaposing the gradient from first-order finite differentiation against the adjoint gradient (Figure \ref{fig1}(c)). Given that the adjoint gradient is exact (assuming its correct computation), one anticipates the discrepancy between the two gradients to diminish as $\delta p$ reduces. However, at a certain threshold of $\delta p$, floating point errors start to dominate over a theoretically improved finite differentiation, rendering the gradient from finite differentiation increasingly inaccurate. Consequently, a hockey stick shaped divergence between the numerical and adjoint gradients is predicted for diminishing values of $\delta p$ \cite{fichtner2021lecture}. Figure \ref{fig1}(c) shows that the gradient test indeed behaves as expected.


\subsection{Algorithm complexity}\label{s:complexity analysis} 

For computing a Berry phase with $I$ discrete time or Bloch wavenumber points, the algorithmic complexity of the adjoint method corresponds to the sum of  $I$ eigenvalue solver instances plus $I$ adjoint problems. The latter equates to $I$ system of equation solutions. Significantly, this complexity is not influenced by the number of model parameters. On the other hand, $O$-order finite differentiation necessitates $O\times I$ eigenvalue problem solutions for each model parameter, leading to a complexity proportional to  $O\times I \times (N_p+1)$ eigenvalue solutions. Thus, as $N_p$, the number of model parameters grows, the adjoint method outperforms finite differentiation. Additionally, in contrast to finite differentiation which suffers from truncation and floating point errors as $\delta p$ approaches $0$, the adjoint method provides an exact gradient within the approximation of the forward discretization.

\section{Generalization to Multiband Berry Phases} \label{s:Multiband}

In this section, we discuss the generalization of the proposed method to the multiband, or non-Abelian Berry phase, which is computed using the Wilson loop approach~\cite{wilson1974confinement}. For a detailed physical derivation and the relation to polarization, maximally localized Wannier functions, or topology, we refer to, e.g., \cite{vanderbilt2018berry, wilczek2012introduction, neupert2018topological, wang2019band}. A discussion on the appearance of these quantities in classical mechanics and how they differ from the quantum case can be found in~\cite{bosch2023differences}.

We consider a set of $B$ isolated bands below a band gap of interest and introduce the overlap matrix as
\begin{multline}
    \mathbf{U}^{(i,i+1)}_{n,m} = \Braket{\mathbf{n}_i|\mathbf{M}|\mathbf{m}_{i+1}} 
    \\ = \Braket{\hat{\mathbf{n}}_i|\hat{\mathbf{M}}|\hat{\mathbf{m}}_{i+1}} + i\Braket{\hat{\mathbf{n}}_i|\hat{\mathbf{Q}}\hat{\mathbf{M}}|\hat{\mathbf{m}}_{i+1}}, \\ n,m \in B,  
\end{multline} 
where we have used the definitions in equations ~\eqref{eq: aug state}-\eqref{eq: Q}. We assume a closed loop, i.e., $\mathbf{n}_0 = \mathbf{n}_I$, for all bands. The gauge-invariant multiband Berry phase is given by~\cite{vanderbilt2018berry}
\begin{equation}
    \gamma_{\text{tot}}\big[\{\{\mathbf{n}_i\}_{i=0}^I\}_{n\in B},\mathbf{p}\big] \approx \sum_{i=0}^{I} \text{Im}\ln\det\mathbf{U}^{(i,i+1)}.
\end{equation}
Note that the above expression is exact in the limit $I \rightarrow \infty$~\cite{neupert2018topological, vanderbilt2018berry}. While this would typically require handling points or intervals of degeneracies below the gap, we now argue that this is not needed in practice.

First note that, in the case where all the bands below the band gap of interest are themselves gapped, the multiband Berry phase decomposes into the sum of the Berry phases of the individual bands, i.e.
\begin{equation}
    \gamma_{\text{tot}}\big[\{\{\mathbf{n}_i\}_{i=0}^I\}_{n\in B},\mathbf{p}\big] = \sum_{n \in B} \gamma_n.
\end{equation}
Furthermore, the multiband Berry phase is a smooth function of the system parameters as long as the gap of interest stays open. Thus, an infinitesimal symmetry-breaking perturbation that lifts degeneracies below the gap will only cause a smooth change in the multiband Berry phase, even if individual Berry phases change abruptly. 

Thus, we can always introduce small perturbations to lift all degeneracies without closing the gap of interest, as an infinitesimal perturbation is enough to open a gap but cannot close a finite one. In practice, this can be done by adding a small random Hermitian perturbation, $\delta\mathbf{p}$, to the Hamiltonian or stiffness matrix to break all symmetries in the system. Then the gradient of the multiband Berry phase can readily be computed with the presented method above as
\begin{equation}
    \frac{d\gamma_{\text{tot}}|_{\mathbf{p+\delta\mathbf{p}}}}{dp_m} = \sum_{n\in B}\frac{d\gamma_{n}|_{\mathbf{p+\delta\mathbf{p}}}}{dp_m}.
\end{equation}
If $\delta\mathbf{p}$ is sufficiently small we approximately have that 
\begin{equation}
    \frac{d\gamma_{\text{tot}}|_{\mathbf{p+\delta\mathbf{p}}}}{dp_m} \approx \frac{d\gamma_{\text{tot}}|_{\mathbf{p}}}{dp_m}.
\end{equation}

\section{Conclusions}\label{s:Conclusions}

We developed the adjoint method to compute parameter gradients of the Berry phase. The method can be applied to devices based on adiabatic evolution, as well as to metamaterial systems, where the Berry or Zak phase serves as a topological invariant. The approach is effective for both single and multiband Berry phases. Combined with tools like topology optimization, this method enables the systematic design of continuous space-time materials and devices that exploit geometric and topological wave propagation. 

\backmatter

\bmhead{Acknowledgements}
We wish to thank Frank Schindler, Tena Dubcek, Sebastian Huber, Valerio Peri, Aleksandra Nelson and Tomáš Bzdušek for helpful discussion. Furthermore, C. B. and A. F. acknowledge funding from ETH Zurich. M.S. acknowledges support by ERC grant (Project No. 101040117). Views and opinions expressed are however those of the author(s) only and do not necessarily reflect those of the European Union or the European Research Council Executive Agency. Neither the European Union nor the granting authority can be held responsible for them.

\bmhead{Conflict of interest statement}
On behalf of all authors, the corresponding author states that there is no conflict of interest.

\bmhead{Replication of results}
Source codes to replicated all the results presented in this article can be found in the Supplementary Material. The code will produce Figure \ref{fig1}.

\bibliographystyle{unsrt}
\bibliography{snbibliography}

\end{document}